# "MXenes: THE KEY TO UNLOCKING THE POTENTIAL METAL-ION BATTERIES-REVIEW"

## N. MIRZAIE DIZAJI[1]

[1] *School of Materials Science and Engineering, Azad University Science and Research Branch, Tehran, Iran; School of Electrical and Computer Engineering, Morgan State University, Maryland, US*

## Abstract

Energy storage devices have been a highly demandable research area in both academia and industry in the last three decades. Although these innovations have the potential to be used as dominant electrochemical energy devices, steps should be taken to minimize charging time while maintaining performance, safety, and charger practicality. The most effective impediment could be the multidisciplinary engineering nature of the cell's electrochemical, structural, and thermo-kinetic perspectives, which form rapidly varying intrinsic states during the charging process. Indeed, many charging strategies fail to adapt to such rapid variations and are based on predefined parameters that enforce and aggregate stress on these devices and reduce their lifespan. The invention of 2D materials has opened a new horizon of possibilities for advanced electrochemical energy storage and conversion applications. MXenes have recently emerged as a new candidate for many applications in energy storage, electromagnetic shielding, communication, smart textiles, and even medicine. Since their first discovery in 2011, they have gathered increasingly considerable interest owing to their unique physical, chemical, and mechanical properties, tunable composition, and surface chemistry. The use of MXenes in batteries was the first application explored by Drexel researchers. MXenes offer high rates and high power advantages compared to other electrochemical energy storage materials; however, large-volume commercial use of these materials would be challenging. Even though much scientific literature has been devoted to research on MXenes, extracting scattered information from the plethora number of journal articles is laborious. This review provides a comprehensive and in-depth overview of the research progress, synthesis, scientific challenges, and design strategies of the new MXene systems, focusing on the influences that MXenes have on the fast-charging performance of metal-ion batteries.

## 1- Introduction

### 1-1-Literature Review

According to the United States (US) Environmental Protection Agency (EPA), transportation takes the largest share (27%) of all economic sectors in terms of greenhouse gas (GHG) emissions [1]. In recent years, efforts to mitigate the effects of this ecological menace have driven rapid progress in the development of energy storage devices [2-5]. Among these, novel electrochemical energy devices play a pivotal role in the future of sustainable energy and are considered to be a key feature in the successful expansion and use of e-transportation ranging from battery-powered electric bikes [6] to e-buses [7], e-trains [8], and EVs [9-11].

Batteries, fuel cells, and electrochemical supercapacitors (SCs) serve as the basis for electrochemical energy storage and conversion devices. Although these three systems rely on electrochemical processes, their charge-storage mechanisms are dissimilar, giving rise to different energy and power densities [12]. Common features are that the energy-providing processes take place at the phase boundary of the electrode/electrolyte interface and that electron and ion transport are separated. It is worth noting that batteries, fuel cells, and supercapacitors all consist of two electrodes in contact with an electrolyte solution.

In batteries and fuel cells, electrical energy generates through redox reactions that occur between electrodes and electrolytes. As reactions at the anode usually take place at lower potentials than at the cathode, the term negative and positive electrodes are used. The anode oxidizes as a current flows through the circuit, and in turn, there is a reduction at the cathode. The distinction between batteries and fuel cells is based on where energy storage and conversion occur. Batteries are self-contained systems in which the anode and cathode serve as the charge transfer medium and participate as "active masses" in the redox reaction. In other words, energy storage and conversion occur in the same compartment. Fuel cells are open systems where the anode and cathode are just charge-transfer media, and the active masses undergo the redox from the air, or a tank, for instance, fuels such as hydrogen and hydrocarbons. Energy storage (in the tank) and energy storage conversion (in the fuel cell) are thus locally separated.

In electrochemical supercapacitors, energy may not be delivered via redox reactions. A supercapacitor stores potential energy electrostatically. This is what allows the capacitor to rapidly deliver and accept a charge, in addition to tolerating exponentially more change cycles. By the orientation of electrolyte ions at the electrolyte/electrolyte interface, so-called electrical double layers (EDLs) [13] are formed and released, which results in a parallel movement of electrons in the external wire, that is, in the energy-delivering process.

Since the invention of next-generation electrochemical energy devices (EEDs), range anxiety and the length of time required to recharge the batteries, as well as moderate energy density have been an urgent concern. As the high performance of energy storage devices has been limited by sluggish charge carrier transport, many solutions have dispersed in various directions seeking robust model-based charging optimization strategies [14-15]. The behavior of cells and packs subjected to fast charging depends on a multitude of factors spanning multiple scales from atomic to the system level, as illustrated in Figure 1 for lithium-ion batteries [16]. These factors could be categorized into two main groups: charging strategies [17-24] to improve charging performance, and new materials [25-31] to gain fast charging capabilities, high energy, and chemical-structural advancements in EED elements (electrode, electrolyte, separator). These strives have faced issues in ultimate electrochemical energy device performance, lifetime, and stability that hinder the sustainability of EED electrifications, and intensify the need for a pathway out of the barrier to boot.

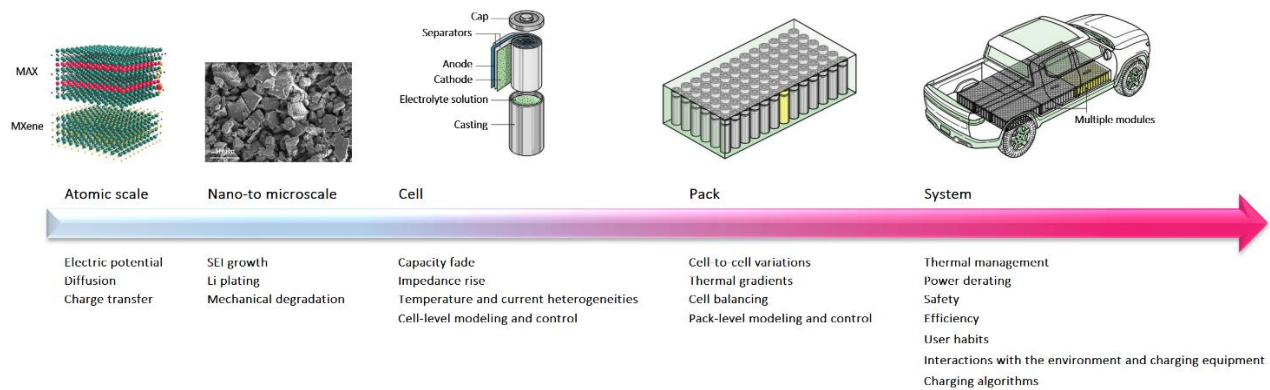

Figure 1. Principle factors affecting lithium-ion batteries fast charging performance at different scales [16,32]

One of the most difficult challenges in cell design is to choose compatible electrode and electrolyte chemistries that provide high capacities and can work effectively at a high rate. In the case of Li-ion batteries, considerable research has been undertaken to create anode materials that enable quick charging while avoiding the growth of lithium dendrites. Some well-known anode materials, such as the carbon-based alternatives (graphite, carbon nanotubes, diamonds, graphene, or graphene oxide) [33-38], metal-based composites (TiO$_2$ [39], lithium titanium oxide (LTO) [40-41], MnO [42-44]) and alloy composites (such as Si [45-46] and Sn [47]-based compounds) [48], have been investigated with some success.

Meanwhile, 2D materials have been considerably investigated as promising anode materials due to their high surface-to-mass ratio and unique physical and chemical properties, leading to shortened Li$^+$ diffusion pathways, fast electron transport, and an increased number of sites for ion storage. After the discovery of graphene in 2004 by Geim and Novoselov [49], other 2D nanomaterials were also explored and fabricated. Such 2D anode materials mainly include transition metal oxides (TMOs, such as $Co_3O_4$ [50-51], $Fe_3O_4$ [52-54], NiO [55-56], $MnO_2$ [57]), transition metal dichalcogenide (TMDs, such as $MoS_2$, $WS_2$, $MoSe_2$, $WSe_2$, FeS, $FeS_2$, and $CoS_2$) [58] and the transition metal carbide/nitrides (mainly highly conductive MXenes, such as $Ti_3C_2T_x$, $Ti_2CT_x$, $MoTi_2C_2T_x$, $Cr_2TiCZT_x$, $Nb_4C_3T_x$, and $V_2CT_x$) [59-61].

A typical 2D nanomaterial has two dimensions greater than the 100 nm range and one dimension with just a few atomic layers thick (e.g., graphene and MXenes) [62-63]. These 2D materials are crystalline in nature and contain a monolayer of atoms. They exhibit low weight, unique structure, high electrical conductivity, excellent chemical stability, flexibility, and high electron mobility owing to their wide surface area, atomic thickness, and efficient in-plane covalent bond and can be metallic, semiconducting, or insulating. In addition, due to their low production cost and unique features, these 2D materials find great applications in electrochemical energy conversion and storage [64-67], biosensors and bioelectronics [68], topological insulators, optoelectronics [69], catalysis [70], functional membranes for wastewater treatment [71-72], and absorbents [73], and so on (as can be seen in Figure 2 about MXenes.)

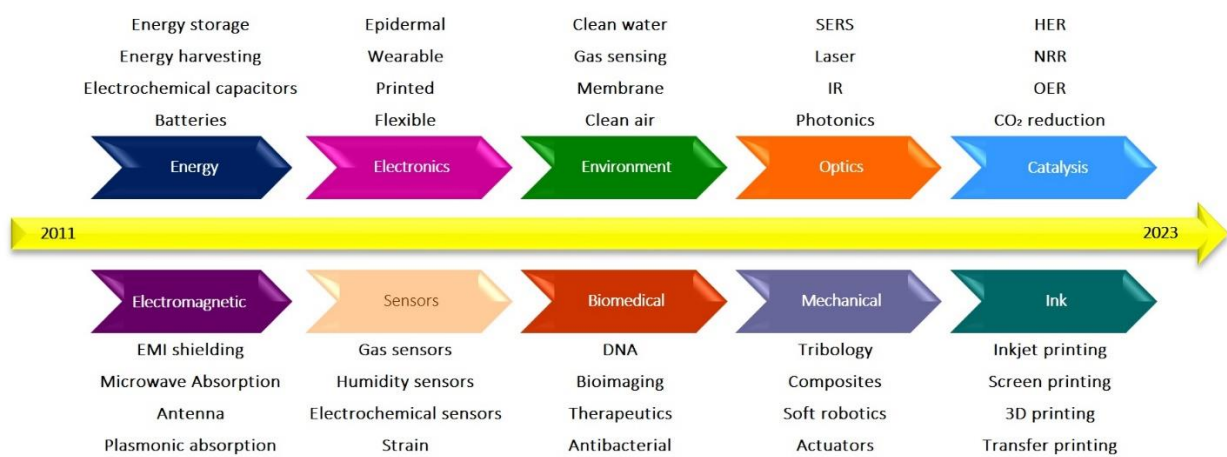

Figure 2. Major applications of MXenes that have been discovered over time

That said, the usage of these materials is beset by challenges in some applications. Take graphene as an example, which exhibits excellent mechanical strength and high mobility charge carriers, is hydrophobic and its usage in electronic devices is impeded by the shortcomings in fabricating larger areas of its sheets. These limitations have led scientists to explore alternative 2D nanomaterials. Among these various vigorously developing areas, MXenes, since their first appearance in 2011, have drawn vast attention due to their intrinsic structure and properties, imparting them particular advantages of being electrode materials in electrochemical energy storage devices [74-81]. MXenes can be derived from the corresponding laminated MAX phases that are governed by the formula $M_{n+1}X_nT_x$ (n = 1-4) [82-84]. These 2D transition metal carbides, nitrides, and carbonitrides possess a host of merits not available in their bulk counterparts: large surface area (due to their atomic thickness), the low energy barrier for electron transportation (because of their crystalline nature), short ion diffusion path, super hydrophilicity, high electrical conductivity (up to 24,000 S cm$^{-1}$ for $Ti_3C_2T_x$) [85-87], flexibility, and the ease of preparation [88-92].

The growing number of publications about MXene-based batteries, fuel cells, and supercapacitors demonstrates their application potential [93-100], making it desirable to provide a comprehensive overview of the development of MXenes and their application in energy storage devices. Therefore, this article aims to systematically provide readers with up-to-date application methods for MXenes in electrochemical energy devices. In this review, we first introduce the unique structures of and synthesis strategies for MXenes. Then, we summarize their recent applications in metal-ion batteries. Finally, we highlight the outlook, pointing out the crucial yet insufficiently explored topics in the fundamental science of MXene materials. We hope that this review article helps advance the energy-storage community and further propels the future development of MXenes for energy conversion and storage.

*1-2-Structures of MXenes*

Being provided with extraordinary electronic, optical, mechanical, and thermal properties, two-dimensional (2D) materials have brought enormous attention from the scientific community because of their wide spectrum of application domains. Since the successful exfoliation of 2D graphene nanosheets from bulk graphite [101], various 2D materials, including transition metals dichalcogenides, boron nitrides, layered double hydroxides, and black phosphorus, have successfully developed. In this regard, MXenes are versatile materials with adjustable and unique structures that have experienced a proportionate expansion in their synthesis methods thanks to their rich surface chemistry during the past decade.

As a class of 2D inorganic compounds, MXenes are the denomination of transition metal carbides, nitrides, or carbonitrides with a thickness of only several atomic layers. MXenes are commonly obtained by chemical delamination of ternary (or quaternary) compounds known as MAX phases as precursors. The MAX phase is the main parent of more than 100 different kinds of metal carbides or/and nitrides, which defined stoichiometry as $M_{n+1}AX_n$, with n = 1, 2, or 3, in which "M" is a d-block early transition metal, and "A" represents main group elements (mainly group 13 and group 14) of the periodic table (e.g. Al, Si, Ge, or Sn) and "X" is carbon, nitrogen, or both [102], as presented in Figure 3. The MAX phase is a 3D crystal structure formed by the bonding and stacking of 2D layered structures. In the MAX phase, the deformed octahedron composed of [$XM_6$] extends laterally in an edge-sharing configuration to form "M-X" layer structures. The "A" layers are located on both sides of the "M-X" structures with metal bonds between the "A" and "M" atoms. M-A bonds and inter-atomic A-A bonds with metallic properties are weaker than M-X bonds with ionic-covalent-metallic contributions, so the 2D layered materials, which are thereafter called MXenes, are produced by removing the "A" layers from the MAX phase. In most cases, a solution containing hydrofluoric acid is used to selectively etch the "A" element (usually Al and Ga) to synthesize MXenes. Therefore, MXenes have a general formula $M_{n+1}X_nT_x$ (n = 1, 2, 3, or 4), where "M" is the early transition metal (such as Sc, Ti, Zr, Hf, V, Nb, Ta, Cr, Mo, and others), "X" is carbon and/or nitrogen, and "T" represents surface terminations (e.g. hydroxyl "-OH", oxygen "-O", fluorine "-F", halogens, chalcogens or their mixtures, denoted as $T_x$) derived from the synthesis procedure [103,84]. MXenes have the representative structures of $M_2XT_x$, $M_3X_2T_x$, and $M_4X_3T_x$, in which the n-layer of the "X" element is covered by the n+1 layer of "M" and abundant surface terminations. Since the first discovery of the $Ti_3C_2T_x$ compound, more than 30 different configurations of MXenes have already been successfully synthesized, and more than 100 stoichiometrical compositions have been predicted theoretically.

The variety of chemistry associated with the element "M" makes MXenes suitable for various applications. For example, Ying et al. pointed out that V-based MXenes as transparent conductors are more than twice as transparent as their Ti-based counterparts, which makes them candidate materials for solar cells [104]. Moreover, Chen et al. suggested that $V_2CT_x$ MXenes are promising anode materials for aqueous energy storage devices, thanks to their intriguing electrochemical performance, especially the ultra-long cycling life [105]. Nb-based MXenes have emerged as proper materials for energy, biomedical, laser, electromagnetic, and microwave shielding applications due to their unique properties, such as excellent conductivity, specificity in metallic properties, ultrafast photonic and optoelectronic characteristics, high photothermal stability, and biocompatibility [106-109]. Both V and Nb-based MXenes demonstrate an excellent capability to handle high charge-discharge rates in Li-ion Batteries [110-112].

On the other hand, based on density functional theory (DFT) calculations, MXenes with different components of "X" show different characteristics. For instance, Zhang et al. claimed that transition metal carbides have larger lattice constants, monolayer thicknesses, and structural stability than their nitride counterparts. In contrast, the latter have active surface chemistry, higher Young's modulus, and electrical conductivity [113].

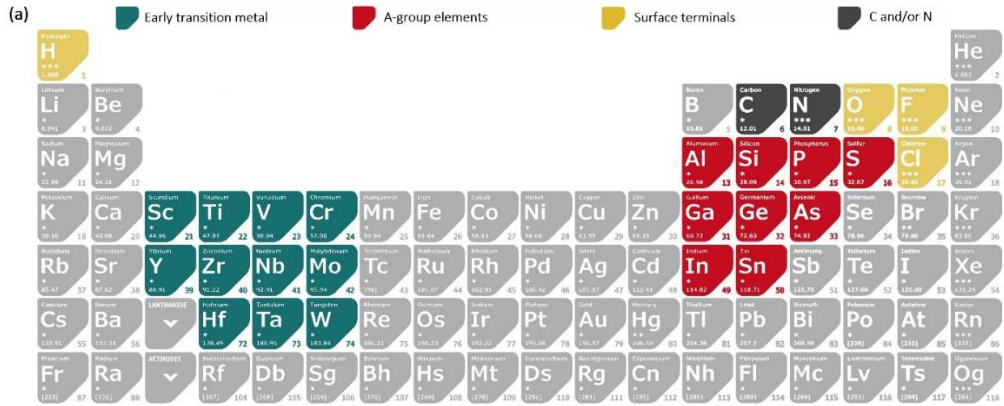
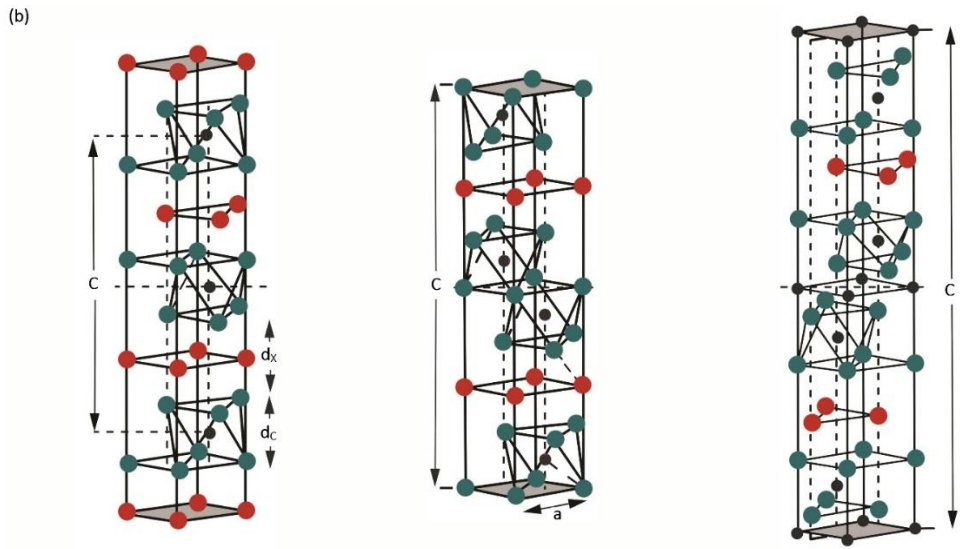
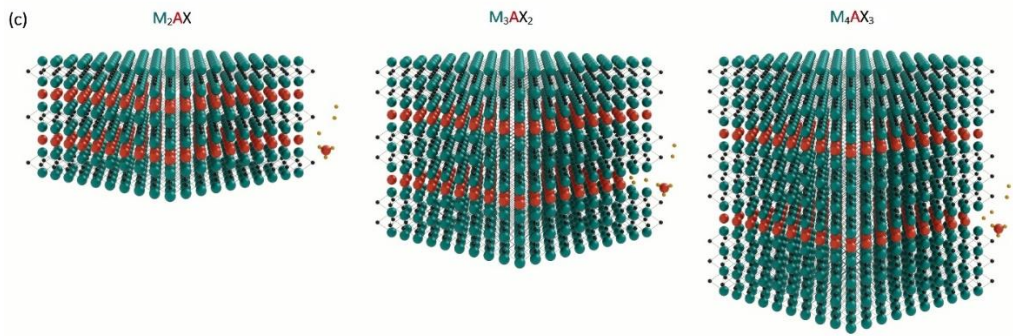
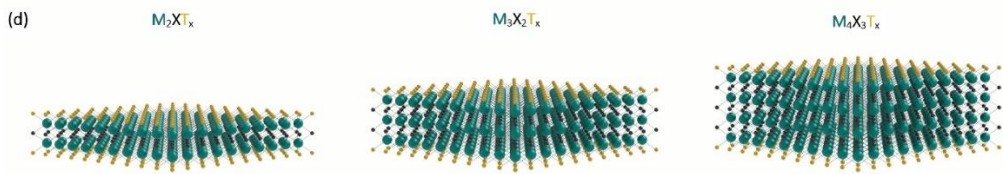

Figure 3. (a) Periodic table showing elements from which MAX and MXene phases are composed, (b) Unit cells of 211, 312, and 413 MAX phases (respectively from left to right.) [91] (c) Atomistic design of MAX phases, (d) Selective etching process of the A-group layers to form MXenes [114]

In addition, the surface terminations, which are introduced during MXene synthesis, affect the adsorption and diffusion behaviors of metal ions on the surface, and the interlayer spacing based on their atomic size, and structural stabilization thanks to saturating the nonbonding valence electrons of transition metal by their low-energy orbitals [115]. As a result, they have a potential impact to control MXenes' electronic [116], mechanical and structural characteristics. Bao et al. mentioned that among terminations, O, S, and N are suitable for the energy storage performance of MXenes, while OH and F are not [117]. Khazaei and his group suggested that without termination, all the MXenes are metallic. However, upon surface functionalization, most $M_2C$ MXenes become semiconductors [118]. This crucial impact makes researchers investigate new materials. For example, Peng et al. represented that the introduction of amino terminated groups on the MXene surface results in excellent electromagnetic interference shielding effectiveness, and improved mechanical properties due to the covalent bonds in the MXene- $NH_2$ polymer composite film [119]. Thomas et al. reported that surface terminations can alter the Fermi level and the associated electron transfer, and therefore, they can play an important role in the work function of MXenes. He claimed that the presence of localized electron states at and around the Fermi level, and subsequently the creation of a high charge density near it, can result in relatively high quantum capacitance [120].

## 2- Synthesis procedure

*2-1-Synthetic strategies for MXenes*

MXenes' distinctive properties derive from the unusual combination of ceramic and metallic behaviors: like ceramics, MAX phases present low density, high hardness, and excellent corrosion resistance; on the other hand, similar to metallic materials, they have high thermal and electrical conductivities and good machinability. These outstanding properties result from the primary bonds: while M-X bonds have a mix of ionic, metallic, and covalent, M-A ones are purely metallic. As opposed to other 2D or 3D materials, such as graphene and transition metal dichalcogenides, that are linked via weak Van der Waals interactions, MAX phases are held together by robust bonds that make them resistant to cleavage through mechanical methods such as shearing. Thus, for the first time, the chemical exfoliation approach made possible the synthesis of 2D materials (MXenes) from layered solids with strong primary bonds (MAX phases). Although MXenes show superior features, their performance may be influenced by external factors such as MAX materials, MXene synthesis conditions, surface functionalities, post-etching treatment, sonication, storage environment, and defect density. As a result, it is crucial to be careful to control factors that maintain the high capability of MXenes for dedicated applications.

So far, two main synthetic strategies for 2D MXenes have been reported: top-down and bottom-up. The top-down strategy refers to the exfoliation of bulk precursors by selectively etching the MAX phase to produce monolayer or multilayer films [121]. The used etching solution usually consists of fluoride ions, such as hydrofluoric acid (HF), ammonium hydrogen fluoride ($NH_4HF_2$), or a mixture of hydrochloric acid (HCl) and lithium fluoride (LiF). A typical example is selectively removing the Al atoms in $Ti_3AlC_2$ to form $Ti_3C_2T_x$, with the aid of aqueous hydrofluoric acid (HF), at room temperature. The latter strategy involves methods, such as chemical vapor depositions (CVD) [122], which show a movement from atoms to a layered structure. Figure 4 depicts the timeline of the evolution of some MXene synthesis strategies, which truly witnesses significant advances in this field.

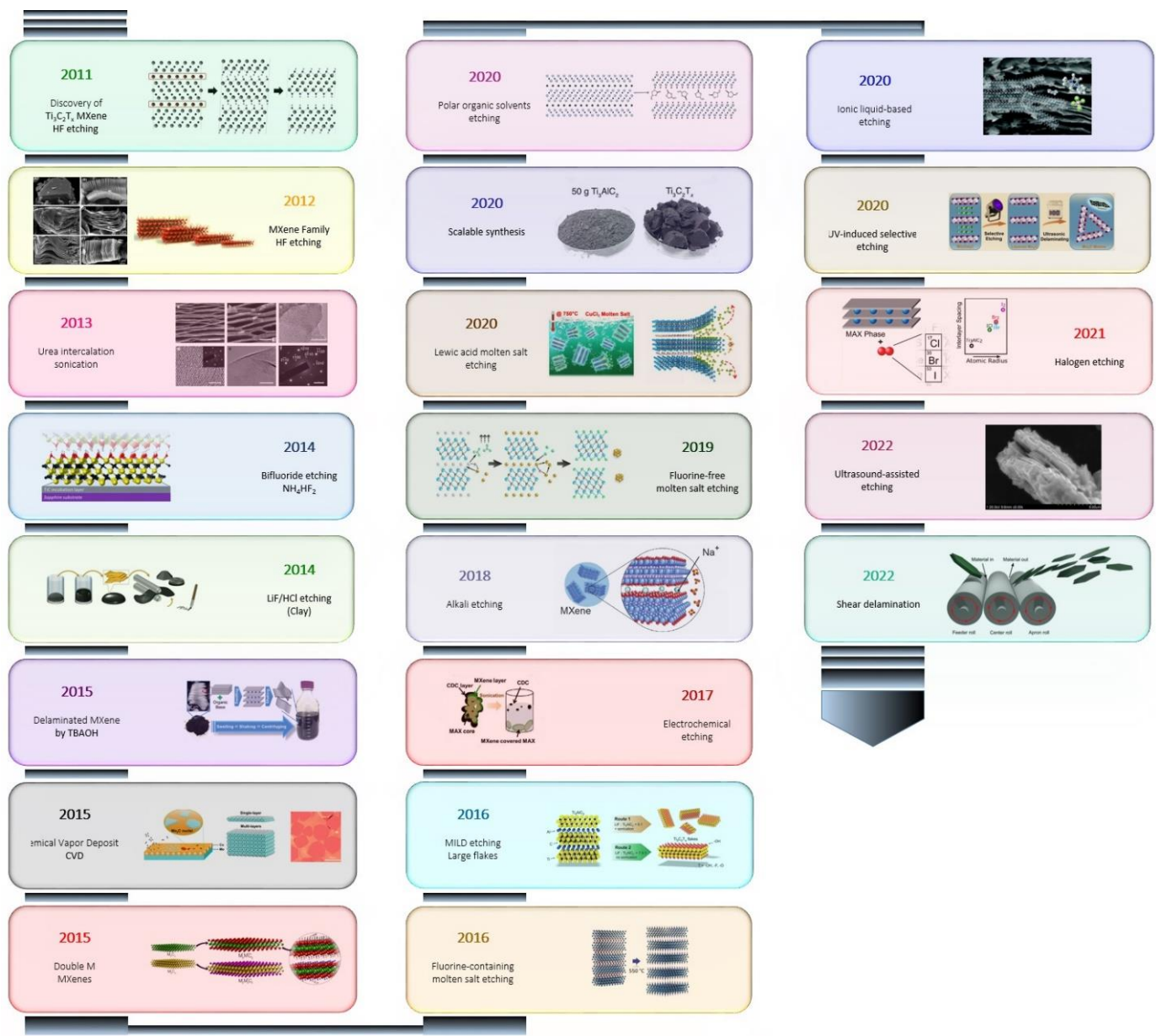

Figure 4. Timeline of the progress in MXene synthesis routes from its first discovery in 2011 [80] until now. In 2012, the synthesis of MXenes was expanded by the exfoliation of immersed MAX phases in HF at room temperature [123]. In 2013, the intercalation and delamination of MXenes by a variety of organic molecules were reported [124]. In 2014, MXenes were fabricated by bifluoride salts [125] and HCl+LiF etchants (clay method) [126]. In 2015, the large-scale delamination of the MXene layers was introduced, using TBAOH [127]. In the same year, the bottom-up CVD synthesis method [128-129], and the existence of two new families of 2D-ordered MXenes were discovered [130]. In 2016, the first two-dimensional transition metal nitride, $Ti_4N_3$-based MXene was synthesized by a fluorine-free molten salt etching method [131]. Later, the MILD route was presented to prepare individual monolayer $Ti_3C_2$ MXene flakes [132]. In 2017, the electrochemical etching of $Ti_2AlC$ to $Ti_2CT_x$ MXene was reported as a green synthetic route [133]. In 2018, the fluorine-free synthesis of high-purity MXenes by an alkali-assisted hydrothermal method was presented [134]. In 2019, fluorine-free molten salt etching was reported to synthesize nanolaminate MXenes [135]. In 2020, Lewis acid molten salt etching [136], polar organic solvent etching [137], ionic liquid-based etching [138-139], and UV-induced selective etching [140] were introduced. Moreover, scalable synthesis made MXenes viable for commercialization [141]. In 2021, halogen etching for MXene fabrication was discovered [142-143]. Then in 2022, new methods such as ultrasound-assisted fabrication of MXenes [144], and mechanical delamination strategies were introduced [145].

*2-1-1- HF wet-chemical etching*

Hydrofluoric acid (HF) is the first reported etching agent to fabricate MXenes. In 2011, researchers from Drexel University, exfoliated MAX phases via immersing $Ti_3AlC_2$ powders in 50% hydrofluoric acid, at room temperature for 2 hours. During the HF etching procedure, the Al layer of the MAX phase reacts with HF (or fluoride ions) to form aluminum fluoride ($AlF_3$) and gaseous hydrogen ($H_2$), which is mainly used to fabricate the MXene phases, as illustrated in the following equations [123]. According to reactions (2), (3), and (4), the surface transition metal (M) reacts with water and fluoride ions to form different surface terminations (-OH, -F and/or -O), depending on whether reaction (2), (3) and/or (4) has next happened.

$M_{n+1}AlX_{n(s)} + 3HF_{(aq)} \rightarrow M_{n+1}X_{n(s)} + AlF_{3(aq)} + 3/2\ H_{2(g)}$ (1)

$M_{n+1}X_{n(s)} + 2H_2O_{(l)} \rightarrow M_{n+1}X_n(OH)_{2(s)} + H_{2(g)}$ (2)

$M_{n+1}X_{n(s)} + 2HF_{(aq)} \rightarrow M_{n+1}X_nF_{2(s)} + H_{2(g)}$ (3)

$M_{n+1}X_{n(s)} + 2\ H_2O_{(l)} \rightarrow M_{n+1}X_nO_{2(s)} + 2H_{2(g)}$ (4)

Apart from the $Ti_3C_2$ reported by Naguib's group [80], other MXenes have also been successfully fabricated by HF etching [146-147]. Meanwhile, HF concentration, etching temperature and time are critical factors in the etching process and are directly related to the performance of MXenes in their applications. According to the concentration and conditions of the HF at room temperature (≥ 10 wt%), most MAXs can be etched with it to generate various accordion-like multilayer MXenes [148], which in turn refers to the exothermic nature of the reaction during the Al removal process and the release of $H_2$. Based on the Alhabeb et al. report, which investigated HF etchant with different concentrations and etching times (5 wt. % for 24 h, 10 wt. % for 5 h, and 30 wt. % for 5 h), low concentrations of HF (5 wt. % for 24 h) would be sufficient for selectively etching aluminum; Although, higher HF concentration could shorten the etching time [149]. In addition, high HF concentrations at a unified reacting time and temperature will increase the proportion of –F terminations [150] and show the fluffiest accordion-like structure. Although, it will represent a smaller interlayer space due to the dissolution of the MAX phase and the formation of ternary fluorides. Thus, etching times would be restricted at high HF concentrations to prevent over-etching. MXene nanosheets terminate with O, OH, F, or more possibly all kinds of combinations, because the etchant solution is rich in fluoride and hydroxyl. Therefore, weak bonds such as hydrogen bonds and Van der Waals bonds replace metal M-Al bonds. However, the bond between the MXene layers is very strong, and the MXene needs to be layered into a single sheet with the help of an intercalator, such as Li-ion or tetrabutylammonium (TBA) ion, etc.

Etching and delamination conditions affect the quality, defects, lateral size, and subsequently the properties and applications of synthesized MXenes [151]. In general, milder etching and delamination conditions are associated with larger MXene flakes with lower defect concentrations that have a wide range of applications in optics, electronics, and electromagnetism [152]. In contrast, high HF concentrations and long etching times can cause small lateral size MXenes with rich active edges that are appropriate for catalysis and gas sensing applications. Besides, MXenes with larger "n" in $M_{n+1}C_nT_x$ require stronger etchant and/or longer etching time. For instance, under the same etching conditions, the etching time required for $Mo_2Ti_2AlC_3$ (n=4) is twice that of $Mo_2TiAlC_2$ (n=3) [151,153].

*2-1-2- In Situ HF-Forming Etching Methods*

To tackle the problem of the corrosive nature of HF etchant, alternative etching routes such as Acid/Fluoride salt etching, Bifluoride Salts Etching, and other in situ HF-Forming systems have also been explored. Since the direct usage of HF could be avoided, in situ HF-Forming methods have simple operation, low energy consumption, and less chemical risk during the etching process, compared with the conventional HF systems. In the in situ HF-forming method, the F ion reacts with the Al atoms of the MAX phase, forming fluoride, $H_2$, and the desired MXenes.

In 2014, Ghidiu et al. reported a new method of producing MXenes by using an HCl/LiF solution for etching the $Ti_3AlC_2$ at 40 °C. The resulting $Ti_3C_2T_x$ could be shaped like clay with strong plasticity and dried into a highly conductive solid or rolled into films tens of micrometers thick using roller pressing [126]. This MXene has superior hydrophilic and film-forming characteristics, as well as high conductivity (an electronic conductivity of 1500 S cm$^{-1}$) [154]. Etching in this system leads to the intercalation of cations and water between MXene layers, expanding the overall interlayer spacing and thus weakening the interaction between layers. Until now diverse research has been done to integrate HCl with different fluoride salts [155-157]; however, among various in situ HF etchants, the HCl/LiF seems to be the most widely used solution that allows the MAX precursor to being easily etched and delaminated just by ultrasound treatment or handshaking.

Halim et al. reported the fabrication of Ti$_3$C$_2$ films by the room temperature selective etching of Al, from sputter-deposited epitaxial Ti$_3$AlC$_2$ thin films in NH$_4$HF$_2$ solution, as an etchant, for the first time [125]. Based on the following equations, the resulting films intercalated with NH$_3$ and NH$_4^+$, enlarging the interlayer spacing and representing higher transparency and resistivity than their HF-etched counterparts.

Ti$_3$AlC$_2$ + 3NH$_4$HF$_2$ → (NH$_4$)$_3$AlF$_6$ + Ti$_3$C$_2$ + 3/2H$_2$ (5)

Ti$_3$C$_2$ + $a$NH$_4$HF$_2$ + $b$H$_2$O → (NH$_3$)$_c$ (NH$_4$)$_d$Ti$_3$C$_2$ (OH)$_x$ F$_y$ (6)

The numerous reports on this subject reflect the possibility of a variety of organic solvents in the presence of NH$_4$HF$_2$ to etch the MAX phases [158-159].

*2-1-3- Electrochemical Etching Methods*

In 2014, Lukatskaya et al. introduced a room-temperature electrochemical etching route to selectively remove the Al atomic layer, using MAX phases as an electrode in dilute HF, HCl, and NaCl solutions as the electrolytes [160]. Since the bonding between the A and M layers is relatively weak, applying the cyclic voltammograms (CVs) between 0 and 2,5 V allows the break of this bond, the removal of the A-layer, and subsequently the electrochemical etching of dense MAX phases. However, the MAX phase will also be over-etched, which causes the removal of the M and A layers to yield carbon-derived carbide (CDC)s or an AX structure. In 2017, Sun et al. reported the electrochemical etching of Al from Ti$_2$AlC electrodes in dilute HCl aqueous electrolyte, eliminating –F terminal groups [133]. The resultant three-layer structure makes of CDC, MXene, and unetched MAX, which will be separated further by bath sonication. The reaction on Ti$_2$AlC as the working electrode is as follows:

Ti$_2$AlC + $y$Cl$^-$ + (2$x$ + $z$) H$_2$O → Ti$_2$C(OH)$_{2x}$Cl$_y$O$_z$ + Al$^{3+}$ + ($x$ + $z$) H$_2$ ↑+ ($y$ + 3) e$^-$ (7)

The reaction on Pt foil as the counter electrode is as follows:

Al$^{3+}$ + 3 e$^-$ → Al (8)

With the further removal of both Ti and Al, the outside layer of MXene (Ti$_2$C(OH)$_{2x}$Cl$_y$O$_z$) is etched into CDC in the presence of terminal groups. Concomitantly, the removal of Ti forms TiO2 on the Pt foil based on the following equation:

Ti$^{4+}$ + 4H2O → Tio2 + 4H+ (9)

Similarly, Song et al. reported the synthesis of 2D niobium carbide MXenes by the fluoride-free e-etching method under 1 V for 4 h at 0.5 M HCl as an electrolyte at 50 °C according to the following [161]:

Nb$_2$AlC + $y$Cl$^-$ + (2$x$ + $z$) H$_2$O → Nb$_2$C(OH)$_{2x}$Cl$_y$O$_z$ + Al$^{3+}$ + ($x$ + $z$) H$_2$ ↑+ ($y$ + 3) e$^-$ (10)

Although the electrochemical etching method offers a green synthetic route with some advantages such as low energy consumption, low reaction temperature, the possibility of recycling the MAX phase as the working electrode several times, and reduced use of corrosive acids, the formation of CDC layer on the surface of the MAX electrode hinders the subsequent etching process and seems not suitable for large-scale MXene preparation.

*2-1-4- Alkali Etching Methods*

Unlike the etching methods mentioned above that all used acids to etch the "A" atom layers; this process presents a two-step etching route, involving the surface treatment of bulk $Ti_3AlC_2$ in NaOH solution for 100 h followed by a hydrothermal treatment via soaking it in $H_2SO_4$ for 2 h at 80 °C. The result would be the selective removal of Al atoms from the layered MAX phase and the formation of a surface exfoliated with the related MXene ($Ti_3C_2$) layers with OH-group-terminations [162]. Although effective, the ultra-high chemical stability of $Ti_3AlC_2$ and the low etching ability of the NaOH solution compared with their HF counterparts restrict the etching of Al atoms just to the surface layer, even in a prolonged period (100 h). In 2018, Li et al. suggested a NaOH-assisted hydrothermal etching technique by the inspiration of the Bayer process, forming the –OH and –O terminated multilayer $Ti_3C_2T_x$ MXene (with 92 wt. % purity) as a result of the selective removal of Al from the MAX phase at 270 °C and in the presence of 27.5 M NaOH [134].

Increasing the alkali concentration and temperature seems to be a successful way to produce more highly hydrophilic F-free MXenes. However, the demerits of using highly concentrated alkali and high temperatures limit the applicability of this method for large-scale MXene preparation.

*2-1-5- Molten Salt Etching Methods*

Although the aqueous etching systems are suitable for a myriad number of Al-containing MAX phases, these methods are not efficient for their non-Al or nitrides counterparts due to two possible reasons. First, the lower cohesive energy of $Ti_{n+1}N_n$ than that of $Ti_{n+1}C_n$ implies its poor structural stability. In addition, the formation energy of $Ti_{n+1}N_n$ is higher than its corresponding carbides thanks to the strong bonding between Ti and Al atoms in the $Ti_{n+1}AlN_n$ structure [163-164]. Being provided with several potential advantages over carbide MXenes, transition metal nitrides are known to be befitting candidates for electrodes in electrochemical capacitors because of higher electronic conductivities. To synthesize the first two-dimensional titanium nitride $Ti_4N_3$, Urbankowski et al. heated the mixture of $Ti_4AlN_3$ powder and fluoride salt (consisting of 59 wt. % KF, 29 wt. % LiF, and 12 wt. % NaF) in 1;1 mass ratio at 550 °C for 30 min under an argon (Ar) atmosphere, following by an extra washing treatment with $H_2SO_4$ and deionized water to dissolve the Al-containing fluorides [131]. Huang et al. reported a general top-down route to prepare Cl-terminated MXenes by etching Zn-based MAX phases in $ZnCl_2$ Lewis acidic molten salt via a replacement reaction between the A-site element of the MAX phase and the late transition-metal halides [135]. Li et al. adopted the Lewis acidic molten salts etching route as a promising way for producing MXenes from various MAX phases with A= Zn, Si, Ga with superior electrochemical performance in high-rate battery and hybrid devices in nonaqueous electrolytes (such as $CuCl_2$, $NiCl_2$, $FeCl_2$) [136]. Khan et al. used this method to etch MAX phases through a direct redox coupling strategy between the "A" element and the cation of Lewis acid molten salt and synthesized fluorine-free $Ti_3C_2$ MXenes with halogen terminals, namely Cl, Br, and I according to the following [165]:

$$Ti_3AlC_2 + 2CuCl_2 \rightarrow Ti_3C_2 + AlCl_4 \uparrow + 2Cu \tag{11}$$

$$Ti_3C_2 + CuCl_2 \rightarrow Ti_3C_2Cl_2 + Cu \tag{12}$$

$$Ti_3AlC_2 + 2.5CuBr_2 \rightarrow Ti_3C_2Br_2 + AlBr_3 \uparrow + 2.5Cu \tag{13}$$

$$Ti_3AlC_2 + 5CuI_2 \rightarrow Ti_3C_2I_2 + AlI_3 \uparrow + 5Cu \tag{14}$$

Since the nature of surface terminations depends on the used molten salt, selecting different types would be a proper way to fine-tune the surface chemistry of desired MXenes. However, in some cases the experimentally prepared MXenes have complex surface group combinations, failing the theoretical principles of MXene models and properties. That is why, despite a wide range of research that has reported the merits of the nonaqueous molten salt method [166-168]; more in-depth investigations are needed to predict the physical and chemical characteristics of produced MXenes.

*2-1-6- Other Etching Methods*

The bottom-up approach has also been reported to prepare large-area defect-free ultrathin MXene crystals, answering the need for potential applications such as electronics and optoelectronics. In 2015, Xu et al. developed a chemical vapor deposition (CVD) method to grow high-quality 2D $Mo_2C$ crystals with a few nanometers of thickness and a large area in the presence of methane and bilayer Cu/Mo foil under temperatures over 1085 °C (the melting point of copper) [129]. In this method, an ultrathin Cu foil is sitting on top of a Mo foil as the growth substrate. High temperatures enable the Cu to be melted and form a Mo-Cu alloy at the liquid Cu/Mo interface. Subsequently, the diffusion of Mo atoms from the interface to the surface of the liquid Cu will cause a reaction between them and carbon atoms, resulting from methane deposition, which makes $Mo_2C$ crystals. The use of very low methane concentration and rapid cooling after CVD growth are fundamental criteria for preparing clean ultrathin crystals [128,169]. Ion sputtering (PVD- physical vapor deposition) is also a proposed route to prepare ultrathin MXene nanosheets. Vacik et al. synthesized the thin film $Ti_2C$ composites by ion beam sputtering of Ti and C (using low energy ion facility (LEIF): 25 keV $Ar^+$ with a current of 400 µA) followed by consecutive thermal annealing under high vacuum [139].

The use of electromagnetic waves to fabricate MXenes from bulk precursors has also been proposed. Mei et al. synthesized free-standing 2D $Mo_2C$ MXene nanosheets from ternary nanolaminate $Mo_2Ga_2C$ via UV-induced selective etching method and following ultrasonic delamination as promising anode materials for both lithium-ion and sodium-ion batteries [140]. In this method, the $Mo_2Ga_2C$ is stirred under UV light (100 W) in the phosphoric acid solution for 3-5 h to obtain layered $Mo_2C$ powder. Although effective in eliminating the use of harmful and highly corrosive acids, the surface of the resultant MXene is not smooth and includes a range of mesopores. In 2022, Zhang et al. suggested an efficient approach to synthesizing $Ti_3C_2T_x$ MXene by using dilute hydrofluoric acid (2 wt. %) under the aid of ultrasound, shortening the etching time to 8 h while exhibiting an outstanding energy storage stability [144].

Halogen has also been introduced as the etchant for MAX etching. Shi et al. reported a fluoride-free, iodine ($I_2$) assisted synthetic process (IE), preparing the –OH and –O terminated $Ti_3C_2T_x$ MXene, rich in oxygen terminal groups. IE involves the iodine etching of the MAX phase in anhydrous acetonitrile ($CH_3CN$) at 100 °C and subsequent delamination in HCl solution to remove the $AlI_3$ generated during the etching process [142]. Jawaid et al. presented a room-temperature etching strategy that utilizes elemental halogen and interhalogens ($Br_2$, $I_2$, ICl, IBr) in anhydrous organic media to synthesize MXenes with homogeneous Cl, Br, or I terminations [143].

In addition, to avoid the negative impacts of using water as the main solvent, an effective route was reported to etch and delamination of MXenes by using various organic polar solvents in the presence of ammonium dihydrogen fluoride, based on the dissociation of $NH_4HF_2$ into $NH_4F$ and HF in polar solvents. To etch the MAX phase, a mixture of $Ti_3AlC_2$, propylene carbonate (PC), and $NH_4HF_2$ is stirred at 500 rpm, inside an Ar-filled glove box for 196 h at 35 °C, which is followed subsequently by a washing and delamination procedure in the presence of PC to exfoliate multilayer MXene [137].

*2-2- The Intercalation and Delamination Strategies of Multilayer MXenes*

Being provided with a high specific surface area, good hydrophilicity, and tunable functional surface, single-layer MXene nanosheets could be reached by intercalation and delamination of stacked MXenes [170-172]. Meanwhile, because of the strong interaction between the MXene layers, the selected way to break the prevailing interlayer forces plays a vital role in successfully separating the stacked accordion-like nanosheets of MXene [173-175].

Based on the similarities of MXenes to other 2D materials such as graphene, and clays [80], in 2013, Mashtalir et al. examined the common organic intercalators such as urea, hydrazine monohydrate ($N_2H_4.H_2O$ as HM), HM dissolved in *N,N*-dimethylfolmamide DMF, and DMSO successfully for the exfoliation of multilayer $Ti_3C_2T_x$, found that the dimethyl sulfoxide (DMSO) is a promising material to delaminate $Ti_3C_2$-based MXenes into single-layer one. This report shows that the interlayer spacing of the multilayer $Ti_3C_2T_x$ was increased from 19.5 ± 0.1 to 35.04 ± 0.02 Å after the intercalation by DMSO molecules, which can significantly reduce the forces between strongly bonded MXene layers and facilitate the subsequent delamination after a weak sonication, forming a stable colloidal solution in water.

In addition, the fourfold capacity of these nanosheets compared to their multilayer counterparts has made them the ideal material for use as anodes in lithium-ion batteries [124]. Since this approach has only been efficient with $Ti_3C_2T_x$, Naguib et al. reported a universal route for the large-scale delamination of MXenes via relatively large organic molecules, the tetrabutylammonium hydroxide (TBAOH), choline hydroxide, and *n*-butylamine. He claimed that the organic treatment would considerably reduce the concentration of –F groups and increase the oxygen content [127]. Han et al. argued that the strong interaction between $Ti_3C_2$ layers due to the high-energy barrier of the Ti-Ti bond and residual Ti-Al bond is the most crucial impediment against the sufficient diffusion of intercalation reagent, resulting in a low yield. Thus, he suggested a hydrothermal-assisted intercalation (140 °C, 24 h) to help the diffusion and intercalation of tetramethylammonium hydroxide (TMAOH) as the reagent in the presence of an antioxidant (ascorbic acid) to boost the yield of 2D MXene sheets while protecting them from oxidation during the HAI process [176]. Similarly, some other research has pointed out the microwave-assisted method to intercalate and delaminate the MXenes thanks to several advantages, namely short processing time and chemical stability [177-178]. Based on the amphoteric nature of Al, Xuan et al. proposed that the TMAOH, which is the main commercial etchant for Al, can act simultaneously as an etchant and intercalator. In this etching-delamination process, TMAOH reacts with the Al atomic layer of the MAX phase, and the resulting $Al(OH)_4^-$ cleavage the Ti-Al bonds, followed by intercalation of the bulky $TMA^+$ cations and delamination of MXenes with $Al(OH)_4^-$ terminations without any external shearing force [179]. After the intercalation stage, the sediment of non-delaminated multilayer MXenes or the stacked nanosheets could be separated through centrifugation, and the supernatant of MXene nanosheets will be collected.

In the case of inorganics, the spontaneous and electrochemical intercalation of cations, including $Li^+$, $Na^+$, $K^+$, $NH_4^+$, $Mg^{2+}$, and $Al^{3+}$, would be enlarged the interlayer space of multilayer MXenes [180-182]. Ghidiu et al. studied the effect of the exchange of intercalated alkali and alkaline earth cations and $H_2O$ between the layers of $Ti_3C_2T_x$, referring to the similarity of MXenes with expanded clay minerals due to the reversible increase in the c lattice parameter with $H_2O$ uptake, and the effect of intercalate cations in surface chemistry and interlayer spacing of MXenes, as well as their subsequent properties. He claimed that the c lattice parameter change and discontinuous structural expansion could be a result of the hydration properties of selected alkali metal ions accommodated between MXene layers [183]. As mentioned in the 2-1-2 section, the HCl/LiF etching method is the promising route that eases the delamination of obtained multilayer MXenes [184-185]. The delaminated monolayer MXene exhibit high stability in some solvents such as dimethyl sulfoxide (DMSO), *N,N*-dimethylformamide (DMF), *N*-methyl-2-pyrrolidone (NMP), propylene carbonate (PC), and ethanol (EtOH) [186]. The resulting uniform colloidal dispersion would be advanced by different approaches, namely wet chemical synthesis [187], fabricating hybrid fibers [188], flexible free-standing films [189-190], and vertical arrays [191].

Some other synthesis improvements have removed the extensive sonication of etched products. The minimally intensive layer delamination (MILD) is one of those methods that increased the size of $Ti_3C_2T_x$ MXene flakes up to several micrometers [132]. In addition, in order to achieve ultra-large $Ti_3C_2T_x$ MXene flakes, Shekhirev et al. proposed the soft delamination approach, eliminating any sonication and shaking that cause the breakage of flakes into smaller particles [192]. Inman et al. presented an alternative approach by shearing multilayer MXenes with a three-roll mill to produce single- and few-layer $Ti_3C_2T_x$ flakes [145]. Mechanical delamination segregates the nanosheets by longitudinal or transverse stress on the surface of layered MXenes. Zhang et al. claimed that the power-focused delamination (PFD) strategy significantly improves the exfoliation efficiency and enhances the yield by 6.4 times compared to sonication exfoliation and MILD methods [193]. The gentle water freezing-and-thawing (FAT) approach also seems to be a simplified and sufficient strategy to exfoliate multilayer-MXenes [194-195].

### 3- Characterization

Owing to the rapid exploration of numerous MXenes, different advanced characterization techniques are necessary to find the suitable MAX precursors, synthesis method, composition, and structure based on the requested final properties.

So far, many methods have been reported to synthesize the precursors of MXenes, such as the high-temperature reaction of a powder mixture in a furnace [196], hot isostatic pressing [197-198], self-propagating high-temperature synthesis [199-200], molten salt synthesis [201-203], spark-plasma sintering [204], microwave synthesis [205-206], magnetron sputtering [207-208], hydride cycle [209], in which high-temperature synthesis is the most common. Since the properties of MXenes are directly influenced by their parents, in the first stage, a wide range of factors should be considered: structural motifs, pure precursors with possible impurities (being assured that there is just one MAX phase in the precursor through X-ray diffraction, and calculation of lattice parameters), nature of the material (investigation of the arrangement or distribution of atoms within the structure, as well as the existence of different MAX polytypes by TEM or PDF), and particle size.

In the next step, to warrant having a successful MXene synthesis process, focusing on the following is essential: color change as a result of converting the MAX phase to the respective MXene, XRD patterns, scanning electron microscopy (SEM) images, PDF (as a valuable way to study surface groups, both M and X site elemental arrangement, and intercalants), X-ray absorption spectroscopy information (the oxidation state of atoms, bond length, and coordination number, as well as surface reactions on MXene sheets), and the lateral flake dimensions and their thickness (by atomic force microscopy) [210].

In order to investigate the chemical composition (especially surface groups) of MXenes, various methods such as transmission electron microscopy/spectroscopy (with EDS analysis) [211], electron energy loss spectroscopy [212-214], X-ray photoelectron spectroscopy (XPS) [215-217], Raman spectroscopy [218-219], and nuclear magnetic resonance [220-221] would be applied.

## 4- Applications of MXenes in Metal-ion Batteries

Since the pioneering work of Drexel University researchers that found the first MXene, these 2D materials have been widely used in energy storage devices such as lithium-ion batteries due to their layered structure and excellent conductivity. Besides, the tunable composition and surface chemistry, mechanical properties, and thermal stability make MXenes promising candidates for a wide range of industrial applications [222], such as electrochemical sensors [223-224], electromagnetic interference (EMI) shielding materials [225-228], inkjet printing [229-230], water treatment [231], catalysis [232-233], biomedical [234-236], corrosion and tribology [237]. Based on the vital position of MXene materials in energy storage and conversion applications, here, MXene and MXene-based composites for metal-ion batteries are summarized.

Electrification of transport using renewable power has been one of the many ways to address the pervasive problem of climate change and respond to the need for clean energy; In the meantime, batteries have played an essential role in these electric vehicles. Li-ion batteries (LIBs) are widely used in energy storage and conversion devices due to their commercial viability. However, there are two primary drawbacks associated with these batteries. Firstly, the scarcity of lithium resources on earth, as discussed in reference [238]. Secondly, the formation of Li dendrites represents a safety and lifetime concern, requiring researchers to explore new rechargeable metal-ion batteries such as Na, K, Mg, Ca, Zn, and Al ions. Sodium is an attractive alternative since it belongs to the same group as lithium and shares similar properties. The main advantage of Na-ion batteries is sustainability, which is crucial for the world striving to be free of carbon-based energy sources [239-240]. In addition, the abundance of sodium and its low cost could be rendered these batteries a promising candidate [241]. Multivalent-ion ones such as Mg, Ca, Zn, and Al-ion batteries [242] have also sparked the immense interest of researchers because of their double or triple amounts of transferred electrons during the electrochemical process compared to their monovalent counterparts, endowing them significantly high energy densities. These alternative batteries also have attracted increasingly more attention due to abundant available resources on nature, high theoretical capacities, and environmentally friendly attributes. On the other hand, most electrode materials suitable for LIBs, such as carbon materials, exhibit poor electro-activity or low capacity for non-lithium-ion batteries [243]. Overall, looking for next-generation electrode materials to accommodate both lithium and non-lithium ions is one of the research hotspots in the current battery field.

Graphite is one of the most used lithium-ion battery anode materials [244]. It has a theoretical capacity of 372 mAh/g [245-248], which cannot satisfy the rapidly growing demand for energy-related applications. Some candidate materials such as silicon, germanium, tin, and carbon nanotubes have emerged with improved capacity compared to graphite electrodes. However, their poor stability will cause volume change and internal deformation of the LIBs during lithiation and delithiation, which seriously influences the safety and life of the battery [249-250]. The global energy crisis and the migration from fossil fuel-driven vehicles to electrically powered ones have led to extensive research on the use of 2D materials, such as exfoliated graphene, graphene nanosheets, reduced graphene oxides, silicone, phosphorene, and borophene, owing to their exceptional physical and chemical properties [251-252]. Since the successful exfoliation of graphene, an increasing number of two-dimensional (2D) materials have aroused enormous research interest. Graphene is considered to be the mother of all graphitic materials like fullerenes, carbon nanotubes, and graphite. It has created a tremendous desire among both physicists and chemists due to its various fascinating properties [253]. However, high hydrophobicity, chemistry, and weak Van der Waals bonding between the graphene layers reduce its energy storage efficiency.

The birth of MXenes by a group of scientists from Drexel University in 2011 [254] opened a new horizon in the use of two-dimensional layered materials in the field of energy storage and conversion and many other applications. The metallic or narrow band gap semiconductor characteristics of MXenes, the type of interlayer forces, and the high specific area make it a potential candidate for LIBs anode material [255]. The high electrical conductivity of MXenes can accelerate the adsorption and desorption of the charge and enhance the diffusion rate of the electrolyte ions, and their high electrochemical activity would be beneficial to the fast and reversible faradaic redox reaction at the surface for charge storage in electrochemical devices [256-257]. The lithiation and delithiation mechanisms were found to be Li intercalation and deintercalation between the MXene layers [258]. Generally, MXenes with n=1 ($M_2X$) should have higher gravimetric capacities in contrast with their higher-order counterparts, such as $M_3X_2$ or $M_4X_3$, because the former has fewer atomic layers than the latter (3 atomic layers vs 5 and 7, respectively). Moreover, $M_2X$-based MXenes should possess higher specific surfaces as compared to higher-order ones [259].

An essential property of MXenes is that they can not only adsorb ions on their surface but the mobility of the ions is affected by the MXene's termination atoms [260]. However, functional groups on the surface of MXenes, such as O, OH, and F tend to reduce the Li storage capacity. On the other side, interlayer collapsing and severe restacking or van der Waals forces between the layers have restricted the access of electrolyte ions and adversely affect intrinsic electrochemical properties. Various effective materials have recently been introduced to combine with MXenes and form MXene hybrid materials to address the problem of restacking aggregation and improve electrochemical performance. The hybrid material has led to high ionic mobility, high diffusivity of $Li^+$-ions, and electrical conductivity apart from specific capacity [261-263].

One of the various strategies is forming van der Waals heterostructures stacked by different types of two-dimensional sheets, which combine the advantages of both materials and stabilize the overall structure. Du et al. investigated the Li adsorption properties of MXene/graphene heterostructures by choosing $Ti_2CX_2$ (X= F, O, and OH) as representative MXenes. He claimed that graphene could effectively separate MXene layers and avoid restacking effects, rendering more electroactive sites accessible in the electrodes. Based on the results, the heterostructures provide enhanced electric conductivity, Li adsorption strength, and mechanical stiffness, while maintaining high Li mobility. They suggested that $Ti_2CO_2$/graphene heterostructure (with the highest specific capacity at 426 mAh/g) is the most promising electrode material for high-performance MXene-based Li-ion batteries [248].

Aierken et al. studied the Li atom intercalation in MXene ($M_2CX_2$ where M= Sc, Ti, V, and X= OH, O)/graphene vertical bilayer heterostructures for Li battery applications, considering the less stability, lower capacities, and very high Li diffusion barriers of F terminated MXenes as compared to their O counterparts. They pointed out that although the binding of a Li atom on mono-$M_2C(OH)_2$ was not favorable due to the Coulomb repulsion between positively charged Li ions and H atoms, Li absorption of the bi-$M_2C(OH)_2$ + Li was made stable again through the van der Waals interaction that competes with foregoing repulsive interactions. They also confirmed that the binding energy of the Li atom on the O-terminated MXene monolayer is stronger than that on the OH-terminated ones. He reported that all the systems related to $Sc_2CO_2$ experienced severe structural distortion upon Li interactions, and claimed that this type of system was unstable for Li intercalation. In the case of $M_2CO_2$ + Gr + Li, $E_b^{Li}$ had slightly decreased for M= Ti and V thanks to the reduced van der Waals interactions concerning the pristine bilayers due to the increase in interlayer separation. Nevertheless, they identified three promising candidates with a large Li atom adsorption energy, namely $Sc_2C(OH)_2$ + Gr + Li, $Ti_2CO_2$ + Gr + Li, and $V_2CO_2$ + Gr + Li. In addition, They found that $Sc_2C(OH)_2$ + Gr is unstable against Li loading at higher concentrations; though, all the possible Li absorption sites can be occupied without compromising stability in $Ti_2CO_2$ + Gr and $V_2CO_2$ + Gr, leading to an average open circuit voltage of 1.49 V for the former and 1.93 V for the latter. These findings suggest that $V_2CO_2$ when exposed to Li undergoes a reversible structural transformation, but this transformation is halted by graphene. The study also concluded that the heterostructures of $Ti_2CO_2$ + Gr + Li and $V_2CO_2$ + Gr + Li are potential contenders for lithium-ion battery applications. Especially, Aierken et al. pointed out that $Ti_2CO_2$ - Gr bilayer has the highest gravimetric capacity due to its light formula unit, also the $Ti_2CO_2$ MXene offers a compromise between capacity and kinetics since the calculated diffusion barriers were the lowest among the other considered systems and were lower than that of graphene [264].

Zhao et al. studied the adsorption and diffusion of Li in a series of different heterostructures composed of MXenes ($M_2CO_2$ where M= Sc, Ti, V) and B-doped graphene (BDG). Their results showed that the M_Sc/BDG heterostructure represented the highest electron transfer between the interfaces, the largest change in its electronic structure, and the strongest adsorption of Li. They proposed that the M_Sc-based heterostructure with the highest theoretical capacitance (472 mAh/g) compared to the M_Ti-based (439-404 mAh/g) and M_V-based (378-380 mAh/g) heterostructures is a suitable candidate for an efficient anode material for LIBs [265].

Transition metal dichalcogenides (TMDs) such as $MoS_2$, $WS_2$, and $MoSe_2$ nanosheets have been uniformly interspersed on the surface of MXenes to mitigate the restacking problem of the MXenes and also amend the electrochemical redox kinetics due to the synergistic interaction between MXenes and TMDs. The MXene-TMD composites have demonstrated excellent reversible specific capacitance, superior coulombic efficiency (CE), and improved cyclability and rate performance in electrochemical applications [266]. Yuan et al. designed the heterostructures of $MoS_2/M_2CS_2$ (M= Ti, V) as electrode materials for metal-ion batteries and argued that the two heterostructures show stable structure and metallicity. The low diffusion barriers benefit from the S functional group of $M_2CS_2$ and imply the excellent rate performance in all batteries under review. The capacities for Na or Li-ion were twice higher as those for K or Mg ion, while the average open-circuit voltage for NIBs is all much lower than those for LIBs. Therefore, the higher capacities and lower voltages further suggest that $MoS_2/M_2CS_2$ are more promising candidates for NIBs than LIBs [267]. Li et al. studied the structural, electrical, and chemical performances of two heterostructures $VS_2/Ti_2CT_2$ (T= O and S) as anode materials for metal-ion batteries. Their results showed that compared to graphene/ $Ti_2CO_2$ and $MoS_2$/ $Ti_2CO_2$ heterostructures, $VS_2/Ti_2CT_2$ simultaneously possessed metal conductivities and stronger adsorption strength towards Li, Na, and Mg atoms. Li et al. presented that both $VS_2/Ti_2CO_2$ and $VS_2/Ti_2CS_2$ are promising high-performance anode materials for Mg-ion batteries, and $VS_2/Ti_2CS_2$ is a preferred anode material for Na-ion batteries [268].

Wen Xi et al. have studied MXene/transition metal oxide heterostructures and suggested them as promising electrode materials for energy storage devices thanks to the synergistic effect of conductive MXenes and active TMOs. They claimed that the hierarchical MXene/TMO heterostructures not only prevent the stacking problem of MXene sheets but also improve the electronic conductivity and buffer the volume change of TMOs during the electrochemical reaction process [269].

Wei et al. applied MXene/organics (MXene@PTCDA) heterostructures in lithium-ion and sodium-ion batteries and significantly enhanced rate capability and cycling performance than bare PTCDA [270]. A.Papadopoulou et al. showed that adsorption energy depends on the electronegativity of the termination atoms, as well as the distance between the terminations, the ions, and the atoms of the top layer of the MXene. The greater the ionic character of the bond between the termination atoms and the ions, the stronger the bond is, which potentially causes higher capacities for the battery, even in slower charge/discharge rates, depending on the size of the ion [271]. Elsewhere, A.Papadopoulou et al. also studied the adsorption and mobility of a series of ions (Li, K, Mg, and Zn) on the surface of the $Zr_2CS_2$ MXene as the anode electrode, taking advantage of the fact that the S termination atoms lower the diffusion barrier for the moving ion. The results revealed this trend, larger ions result in more hybridization when compared to the $Zr_2CS_2$ MXene structure. Apart from Zn, which detached from the $Zr_2CS_2$ surface and showed outlier results, the Li, K, and Mg ions were strongly bonded to MXene's surface. They claimed that the K ion showed the highest mobility, and $Zr_2CS_2$-K exhibited a lower VOC than the $Zr_2CS_2$-Li and $Zr_2CS_2$-Mg materials; therefore, suggested KIB as the most promising to substitute LIB, especially after taking into consideration K's low cost, abundant resources and comparable energy density to Li [272].

As can be seen, MXenes have a wide range of chemical and structural diversity and excellent metal conductivity. They can also be combined with other materials to upgrade the desired features, which is the reason why MXenes are considered one of the most valuable materials for the next generation of metal-ion batteries.

## 5- Outlook

MXenes are a novel and ever-expanding class of 2D materials, exhibiting plenty of unrivaled properties, driven by the chemical and structural complexity associated with divergent "M," "X," and surface terminations. The introduction of surface terminals during the synthesis process makes MXenes hydrophilic. Hydrophilicity, high metallic conductivity, and high negative surface charge enable them to be dispersed in water, forming stable colloidal solutions of single-layer flakes or liquid crystal slurries with the rheological behavior of clay with no surfactants or additives. The use of MXenes in batteries was the first application explored; however, their application is not only limited to this topic but is also wide ranging from electronics to biomedical. MXenes are the true answer to the need for reassuring materials capable of improving real-world applications. Their high rate/high power ability and redox reactions of transition metals have been a persuasive solution in developing electrochemical high-performance energy storage devices. Their unique surface chemistry, layered structure, superb photoelectric characteristics, and high carrier density make MXenes a promising candidate for alkaline ion batteries. Despite the considerable progress in MXenes' development, the understanding of this emerging family of materials, mainly beyond $Ti_3C_2T_x$, is still at an early stage.

Since cutting-edge industrial technologies are seeking large amounts of cheap but high-quality materials, actions should be taken to mitigate the problems of cost, safety, environmental protection, and stability of MXenes. Although common MXenes are produced from abundant and potentially low-cost materials, most of the production cost is allocated to the synthesis process of both MAX phase precursors and MXenes. There has been a wide range of synthetic methods (ranging from direct and indirect HF-based syntheses to electrochemical, alkaline, halogen, and Lewis acid molten salt etching), introducing to gain desired properties (such as high electrical conductivity to reduce interfacial charge transfer resistance, tunable and uniform $T_x$ coverage towards selective adsorption of the analytes, hydrophilicity, electrochemical stability in aqueous electrolyte and solution processability [175]) for a countless number of applications. However, producing large volumes of MXenes for large-scale commercial productions would be challenging. In this regard, understanding the role of each step, including precursor synthesis, etching-exfoliation, and intercalation-delamination is crucial. Although efficient, different synthetic routes suffer from limiting features: being corrosive, toxic, and environmentally detrimental (HF etching), by-product formation which restricts the etching process and reduces the overall MXene yield (electrochemical etching), the need for high temperatures and alkali concentrations (alkali etching), being difficult and costly to use for mass production (bottom-up strategies such as CVD), and the limitation for effective delamination of multilayered MXenes into a single or few-layer thick MXene flakes (molten salt etching). MXenes are also readily oxidized in humid environments, which is a function of temperature and humidity intensity. This oxidation behavior is an obstacle to the stability of MXenes and affects their large-scale synthesis, storage life, shipping, and long-term practical operations. While deoxygenation, removal of water molecules, low temperatures, and manufacturing of large single-crystal MXene flakes with minimal edge areas are some effective routes, the search for new techniques without compromising the MXene optimum application performance should continue.

Significant efforts have been underway to develop high-performance batteries using MXene-based materials as cathodes, separators, and anodes. Delaminated ultrathin MXene nanosheets are prone to restacking through van der Waals interactions between the layers, further alleviating their surface area and functional groups, which is not conducive to excluding the shuttle effect, improving the sluggish kinetics, and achieving high loadings in metal-ion batteries. However, MXene hybrid materials are a practical answer to avoid MXene restacking and improve the lithium-ion diffusion through the cathode, as well as the electrochemical redox kinetics.

To sum up, research in MXenes is still at a primitive step. It seems that there is still a long way to go before MXenes can be applied in commercial energy storage devices. Thus, we urgently need to boost our understanding of the synthesis routes, properties, and mechanisms of MXenes as electrode materials. It is essential to follow a targeted direction in the ongoing broad studies in MXenes and be supported by more commercial companies for mass production and consumption. Despite the possible challenges, MXenes in high-capacity, high-stability, and fast charge-discharge energy storage systems have excellent prospects for future applications.